\documentstyle[aps,psfig,pra]{revtex}
\newcommand{\ket}[1]{|{#1}\rangle}
\newcommand{\bra}[1]{\langle {#1}|}

\begin{document}

\draft

\title{Entanglement induced by spontaneous emission in spatially
extended two-atom systems}
\author{Z. Ficek$^{1}$ and R. Tana\'s$^{2}$}
\address{$^{1}$ Department of Physics,
The University of Queensland, Brisbane, QLD 4072, Australia
\\ [8pt]
$^{2}$ Nonlinear Optics Division, Institute of Physics, Adam Mickiewicz
University, Pozna\'n, Poland}

\date{\today}

\maketitle

\begin{abstract}
We investigate the role of the collective antisymmetric
state in entanglement creation by spontaneous emission in a system of
two non-overlapping two-level atoms. We calculate and illustrate graphically
populations of the collective atomic states and the Wootters entanglement
measure (concurrence) for two sets of initial atomic conditions.
Our calculations include the dipole-dipole interaction and a
spatial separation between the atoms that the antisymmetric state of
the system is included throughout even for small interatomic
separations. It is shown that spontaneous
emission can lead to a transient entanglement between the atoms even
if the atoms were prepared initially in an unentangled state.
We find that the ability of spontaneous emission to create the
transient entanglement relies on the absence of population in the
collective symmetric state of the system. For the initial state of only
one atom excited, the entanglement builds up
rapidly in time and reaches a maximum for the parameter values
corresponding roughly to zero population in the symmetric state. On
the other hand, for the initial condition of both atoms excited,
the atoms remain unentangled until the symmetric state is 
depopulated. A simple physical interpretation of these results is given 
in terms of the diagonal states of the density matrix of the system.
We also study entanglement creation in a system of two non-identical
atoms of different transition frequencies. It is found that the
entanglement between the atoms can be enhanced compared to that for
identical atoms, and can decay with two different time scales
resulting from the coherent transfer of the population from the
symmetric to the antisymmetric state. In addition, we find that a 
decaying initial entanglement between the atoms can display a revival
behaviour.
\end{abstract}

\pacs{32.80.-t, 32.80.Qk, 42.50.Gy, 42.50.Ar}

%\date{Received \phantom{August 5, 2002}}

\section{Introduction}

The subject of generation of entangled states  has attracted much
interest since it became clear that entanglement lies at the heart of
many new applications which come under the general heading of quantum
information and quantum computation. Several methods of creating
entanglement have been proposed involving trapped and cooled ions or
neutral atoms~\cite{bbtk,koz,hal,flei,sm01,b2,nat1,nat3}. Of particular
interest is generation of
entangled states in two-atom systems, since it is generally believed
that entanglement of only two microscopic quantum systems (two qubits)
is essential to implement quantum protocols such as quantum computation.
It has been shown that entangled states in a two-atom system can be
created by a continuous driving of the atoms with a coherent
or chaotic thermal field~\cite{sm01,afs,klak,zsl}, or by spontaneous 
emission from two distant atoms initially prepared in a coherent 
superposition state or in an entangled state~\cite{phbk,b1,cabr,bgz}. 
In particular, Cabrillo {\it et al.}~\cite{cabr} demonstrated that 
two three-level atoms initially prepared by a weak pulse in an 
entangled state can evolve under spontaneous emission into an 
entangled state of the ground states of the atoms.

The effect of spontaneous emission on entanglement creation 
has also been discussed by others~\cite{gy1,gy2,bash,jak}. 
These studies, however, have been limited to the small sample 
(Dicke) model~\cite{dic}. The disadvantage of the Dicke model is 
that it does not
include the dipole-dipole interaction among the atoms and does not
correspond to realistic experimental situations of atoms located
(trapped) at different positions. In fact, the model corresponds to a
very specific geometrical configuration of the atoms confined to a
volume much smaller compared with the atomic resonant wavelength (the
small-sample model). The present atom trapping and cooling
techniques can trap two atoms at distances of order of a resonant
wavelength~\cite{eich,deb,tos}, which makes questionable the
applicability of the Dicke model to physical systems.

In this paper we study what may be termed "spontaneously induced
transient entanglement" in a system of two interacting atoms. 
In related works, Kim {\it et al.}~\cite{klak} and Zhou {\it et 
al.}~\cite{zsl} have shown that a 
transient entanglement can be created in the Dicke model driven by a 
thermal (chaotic) field. S\o rensen {\it et al.}~\cite{sdcz} have 
proposed a method to produce a transient entanglement in 
Bose-Einstein condensate excited by a single pulse. Turchette {\it et 
al.}~\cite{tur} have recently realised experimentally a transient 
entanglement in two trapped ions. Unlike previous 
works~\cite{klak,zsl,phbk,b1,cabr,bgz,sdcz}, we consider entanglement
creation by spontaneous emission from {\it initially uncorrelated} atoms 
and without the presence of external coherent or incoherent fields. 
We are particularly 
interested in three aspects of entanglement creation by spontaneous 
emission: (1) The dependence of the entanglement creation on specific 
arrangements of initial uncorrelated states, (2) the role of the 
antisymmetric state in the entanglement creation, and (3) sharing and 
transfer of entanglement between two entangled states. We do not make 
the small-sample approximation, so that our results are valid for
arbitrary interatomic separations. The antisymmetric state and its
dynamics are not neglected, even when dealing with small interatomic
separations. We show that spontaneous emission from two spatially 
separated atoms can lead to a transient entanglement of initially 
unentangled atoms. This result contrasts the Dicke model where 
spontaneous emission cannot produce entanglement from initially 
unentangled atoms~\cite{klak,bash}. Moreover, we show that the 
entanglement creation relies crucially on the population distribution 
between the entangled symmetric and antisymmetric states and attains 
maximal values when the population of the symmetric state becomes zero.
This is a rather surprising prediction, since the symmetric state is
an example of maximally entangled state and one might conclude that
its participation in the atomic dynamics would enhance entanglement.

\section{Collective two-atom systems}

We consider a system of two non-overlapping two-level atoms with
ground states $\left|g_{i}\right\rangle$ and excited states
$\left|e_{i}\right\rangle \ (i=1,2)$ connected by dipole transition
moments $\vec{\mu}_{i}$. The atoms are
located at fixed positions $\vec{r}_{1}$ and $\vec{r}_{2}$
and coupled to all modes of the electromagnetic field, which we assume
are in the vacuum state. We consider spontaneous emission from identical
as well as non-identical atoms prepared in two different initial states.
In the case of nonidentical atoms, we assume that atoms have equal dipole
moments $\vec{\mu}_{1}=\vec{\mu}_{2}=\vec{\mu}$, but
different transition frequencies $\omega_{1}$ and
$\omega_{2}$, such that $\omega_{2}-\omega_{1} \ll \omega_{0}=
(\omega_{1}+\omega_{2})/2$, so that the rotating-wave approximation
can be applied to calculate the dynamics of the system.

The time evolution of the system of atoms coupled through the vacuum
field is given by the following master equation~\cite{leh,ag74,ftk87}
\begin{eqnarray}
      \frac{\partial \hat{\rho}}{\partial t} &=&
      -i\sum_{i=1}^{2}\omega_{i}
      \left[S^{z}_{i},\hat{\rho}\right]
      -i\sum_{i\neq j}^{2}\Omega_{ij}
      \left[S^{+}_{i}S^{-}_{j},\hat{\rho}\right] \nonumber \\
      &-& \frac{1}{2}\sum_{i,j=1}^{2}\Gamma _{ij}\left( \hat{\rho}
      S_{i}^{+}S_{j}^{-}+S_{i}^{+}S_{j}^{-}\hat{\rho}
      -2S_{j}^{-}\hat{\rho} S_{i}^{+}\right) \ , \label{eq1}
\end{eqnarray}
where $S_{i}^{+}\ (S_{i}^{-})$ are the dipole raising (lowering)
operators and $S^{z}$ is the energy operator of the $i$th atom. In
Eq.~(\ref{eq1}), $\Gamma_{ij}\ (i=j)$ are the spontaneous
emission rates of the atoms, equal to the Einstein $A$ coefficient
for spontaneous emission, whereas $\Gamma_{ij}$ and $\Omega_{ij}\
(i\neq j)$ describe the interatomic coupling~\cite{leh,ag74,ftk87},
and are the collective damping and the dipole-dipole interaction
potential defined, respectively, by
\begin{eqnarray}
\Gamma_{ij}=\Gamma_{ji}&=&
\frac{3}{2}\Gamma
\left\{ \left[1 -\left( \bar{\mu}\cdot \bar{r}
_{ij}\right)^{2} \right] \frac{\sin \left( k_{0}r_{ij}\right)
}{k_{0}r_{ij}}\right.  \nonumber \\
&&\left. +\left[ 1 -3\left( \bar{\mu}\cdot
\bar{r}_{ij}\right)^{2} \right] \left[ \frac{\cos \left(
k_{0}r_{ij}\right) }{\left( k_{0}r_{ij}\right) ^{2}}-\frac{\sin \left(
k_{0}r_{ij}\right) }{\left( k_{0}r_{ij}\right) ^{3}}\right] \right\} \ ,
\label{eq2}
\end{eqnarray}
and
\begin{eqnarray}
\Omega_{ij}&=&
\frac{3}{4}\Gamma
\left\{ -\left[1 -\left( \bar{\mu}\cdot \bar{r}
_{ij}\right)^{2} \right] \frac{\cos \left( k_{0}r_{ij}\right)
}{k_{0}r_{ij}}\right.  \nonumber \\
&&\left. +\left[ 1 -3\left( \bar{\mu}\cdot
\bar{r}_{ij}\right)^{2} \right] \left[ \frac{\sin \left(
k_{0}r_{ij}\right) }{\left( k_{0}r_{ij}\right) ^{2}}+\frac{\cos \left(
k_{0}r_{ij}\right) }{\left( k_{0}r_{ij}\right) ^{3}}\right] \right\} \ ,
\label{eq3}
\end{eqnarray}
where $k_{0}=\omega_{0}/c$, $r_{ij} =\left|\vec{r}_{j}-\vec{r}_{i}\right|$
is the distance between the atoms, $\bar{\mu}$ is unit vector
along the atomic transition dipole moments, that we assume are
parallel to each other, and $\bar{r}_{ij}$ is the unit vector along
the interatomic axis.

The master equation~(\ref{eq1}) has been used for many years to study
a wide variety of problems involving the interaction of collective
atomic systems with the radiation field~\cite{ft01}.
Using the master equation~(\ref{eq1}), we can write down the equations
of motion for the components of the density matrix of the two-atom
system in the basis of the product states
$\left|e_{1}\right\rangle \left|e_{2}\right\rangle$,
$\left|e_{1}\right\rangle \left|g_{2}\right\rangle$,
$\left|g_{1}\right\rangle \left|e_{2}\right\rangle$ and
$\left|g_{1}\right\rangle \left|g_{2}\right\rangle$ of the individual
atoms. However, the problem simplifies by working in the basis of the
collective states of the system which contains symmetric and
antisymmetric combinations of the product states. For identical atoms
$(\omega_{1}=\omega_{2})$ the collective states are~\cite{dic,leh}
\begin{eqnarray}
      \left|e\right\rangle &=& \left|e_{1}\right\rangle
      \left|e_{2}\right\rangle \ ,\nonumber \\
      \left|s\right\rangle &=& \frac{1}{\sqrt{2}}
      \left(\left|e_{1}\right\rangle \left|g_{2}\right\rangle
      +\left|g_{1}\right\rangle
      \left|e_{2}\right\rangle\right) \ ,\nonumber \\
      \left|a\right\rangle &=& \frac{1}{\sqrt{2}}
      \left(\left|e_{1}\right\rangle \left|g_{2}\right\rangle
      -\left|g_{1}\right\rangle
      \left|e_{2}\right\rangle\right) \ ,\nonumber \\
      \left|g\right\rangle &=& \left|g_{1}\right\rangle
      \left|g_{2}\right\rangle \ .\label{eq4}
\end{eqnarray}
In the collective state representation, the two-atom system behaves
as a single four-level system with the ground state
$\left|g\right\rangle$, the upper state $\left|e\right\rangle$, and
two intermediate states: the symmetric $\left|s\right\rangle$ and
antisymmetric $\left|a\right\rangle$ states. The most important
property of the collective states is that the symmetric and
antisymmetric states are maximally entangled states. The states are
linear superpositions of the product states which cannot be separated
into product states of the individual atoms.

For non-identical atoms, the collective states of the system contain
non-maximally entangled states, which can be written as linear
combinations of the maximally entangled states
\begin{eqnarray}
      \left|e\right\rangle &=& \left|e_{1}\right\rangle
      \left|e_{2}\right\rangle \ ,\nonumber \\
      \left|s^{\prime}\right\rangle &=& \frac{1}{\sqrt{2}}
      \left[\left(\alpha +\beta \right)\left|s\right\rangle
      +\left(\beta -\alpha \right)\left|a\right\rangle
      \right] \ ,\nonumber \\
      \left|a^{\prime}\right\rangle &=& \frac{1}{\sqrt{2}}
      \left[\left(\alpha -\beta \right)\left|s\right\rangle
      +\left(\alpha +\beta \right)\left|a\right\rangle
      \right] \ ,\nonumber \\
      \left|g\right\rangle &=& \left|g_{1}\right\rangle
      \left|g_{2}\right\rangle \ ,\label{eq5}
\end{eqnarray}
where $\alpha = d/\sqrt{d^{2}+\Omega_{12}^{2}},
\beta = \Omega_{12}/\sqrt{d^{2}+\Omega_{12}^{2}}$, $d = \Delta
+\sqrt{\Omega_{12}^{2}+\Delta^{2}}$, and $\Delta =(\omega_{2}
-\omega_{1})/2$.

Thus, in both cases of identical or non-identical atoms, we can limit
the considerations to the basis of the collective states~(\ref{eq4}).
In this basis, the density matrix elements satisfy the following set
of simple differential equations
\begin{eqnarray}
      \dot{\rho}_{ee} &=& -2\Gamma \rho_{ee} \ ,\nonumber \\
      \dot{\rho}_{ss} &=& -\left(\Gamma
      +\Gamma_{12}\right)\left(\rho_{ss} -\rho_{ee}\right)
      +i\Delta \left(\rho_{as}-\rho_{sa}\right) \ ,\nonumber \\
      \dot{\rho}_{aa} &=& -\left(\Gamma
      -\Gamma_{12}\right)\left(\rho_{aa} -\rho_{ee}\right)
      -i\Delta \left(\rho_{as}-\rho_{sa}\right) \ ,\nonumber \\
      \dot{\rho}_{as} &=& -\left(\Gamma +2i\Omega_{12}\right)\rho_{as}
      +i\Delta \left(\rho_{ss}-\rho_{aa}\right) \ ,\nonumber \\
      \dot{\rho}_{se} &=& -\left[\frac{1}{2}\left(3\Gamma
      +\Gamma_{12}\right) -i\left(\omega_{0}
      -\Omega_{12}\right)\right]\rho_{se} +i\Delta \rho_{ae}
      \ ,\nonumber \\
      \dot{\rho}_{ae} &=& -\left[\frac{1}{2}\left(3\Gamma
      -\Gamma_{12}\right) -i\left(\omega_{0}
      +\Omega_{12}\right)\right]\rho_{ae} +i\Delta \rho_{se}
      \ ,\nonumber \\
      \dot{\rho}_{gs} &=& -\left[\frac{1}{2}\left(\Gamma
      +\Gamma_{12}\right) -i\left(\omega_{0}
      +\Omega_{12}\right)\right]\rho_{gs} + \left(\Gamma
      +\Gamma_{12}\right)\rho_{se} -i\Delta \rho_{ga}
      \ ,\nonumber \\
      \dot{\rho}_{ga} &=& -\left[\frac{1}{2}\left(\Gamma
      -\Gamma_{12}\right) -i\left(\omega_{0}
      -\Omega_{12}\right)\right]\rho_{ga} - \left(\Gamma
      -\Gamma_{12}\right)\rho_{ae} -i\Delta \rho_{gs}
      \ ,\nonumber \\
      \dot{\rho}_{eg} &=& -\left(\Gamma +2i\omega_{0}\right)\rho_{eg}
      \ .\label{eq6}
\end{eqnarray}

Equations~(\ref{eq6}) show that all transitions rates to and from the
symmetric state are equal to $(\Gamma +\Gamma_{12})$. On the other
hand, all transitions rates to and from the antisymmetric state are
equal to $(\Gamma -\Gamma_{12})$. Thus, the symmetric state decays
with an enhanced (superradiant) rate, while the antisymmetric state
decays with a reduced (subradiant) state. Hence, the population of
the antisymmetric state experiences a variation on a time scale of
order $(\Gamma -\Gamma_{12})^{-1}$, which can lead to interesting
effects not observed in the Dicke model. These effects result from
the fact that the set of equations~(\ref{eq6}) has two different solutions
depending on whether $\Gamma_{12}=\Gamma$ or $\Gamma_{12}\neq
\Gamma$. The case of $\Gamma_{12}=\Gamma$ corresponds to the small
sample (Dicke) model, whereas the case of $\Gamma_{12}\neq \Gamma$
corresponds to spatially extended atomic systems. The existence of
two different solutions of Eq.~(\ref{eq6}) is connected with conservation
of the total spin $S^{2}$, that $S^{2}$ is a constant of motion for the
Dicke model and $S^{2}$ not being a constant of motion for a spatially
extended system of atoms~\cite{ftk81,hsf82}. We can explain it by
expressing the square of the total spin of the two-atom system in
terms of the density matrix elements of the collective system as
\begin{eqnarray}
         S^{2}\left(t\right) = 2 -2\rho_{aa}\left(t\right) \ .\label{eq7}
\end{eqnarray}
It is clear from Eq.~(\ref{eq7}) that $S^{2}$ is conserved only in
the Dicke model, in which the antisymmetric state is ignored. For a
spatially extended system the antisymmetric state participates fully
in the dynamics and $S^{2}$ is not conserved. The Dicke model evolves
between the triplet states $\left|e\right\rangle$, $\left|s\right\rangle$,
and $\left|g\right\rangle$, while the spatially extended two-atom system
evolves between the triplet and the antisymmetric states.

\section{Transient entanglement}

The entanglement creation by spontaneous emission is illustrated most
clearly if one assumes that a system of two atoms decays spontaneously 
from initially unentangled (uncorrelated) states. Several different 
measures have been proposed to identify entanglement between two 
atoms, and we choose the
Wootters entanglement measure~\cite{woo}, the concurrence $C$, 
defined as
\begin{eqnarray}
     C = {\rm max}\left(0, \sqrt{\lambda_{1}} -\sqrt{\lambda_{2}} 
     -\sqrt{\lambda_{3}} -\sqrt{\lambda_{4}}\right) \ ,\label{eq7a}
\end{eqnarray}
where $\lambda_{1},\ldots ,\lambda_{4}$ are the eigenvalues of the 
matrix $\tilde{\rho}= \rho (\sigma_{y}\otimes 
\sigma_{y})\rho^{\ast}(\sigma_{y}\otimes \sigma_{y})$ and 
$\sigma_{y}$ is the Pauli matrix. The range
of the concurrence is from 0 to 1. For unentangled atoms $C=0$ whereas
$C=1$ for the maximally entangled atoms. The concurrence involves 
eigenvalues of the complicated matrix $\tilde{\rho}$ and, in general, 
is difficult to 
calculate analytically. Therefore, for the understanding and 
explanation of the entanglement creation via spontaneous emission, we 
will use the Peres-Horodecki (negativity) measure for 
entanglement~\cite{per,horo}. The negativity criterion is given by 
the quantity
\begin{eqnarray}
     E = {\rm max}\left(0, -2\sum_{i}\mu_{i}\right) \ ,\label{eq7b}
\end{eqnarray}
where the sum is taken over the negative eigenvalues $\mu_{i}$
of the partial transposition of the density matrix $\hat{\rho}$ of the
system. The value $E=1$ corresponds to maximum entanglement between
the atoms whilst $E=0$ describes completely separated atoms.

The two entanglement measures, the concurrence $C$ and negativity $E$, 
give the same results for criteria for entanglement, but they can 
give different results for a degree of entanglement~\cite{munro}.

\subsection{Identical atoms}

We begin with spontaneous emission from two identical atoms 
and consider entanglement creation for two different sets of initial 
atomic conditions at $t=0$. In the first, one of the atoms is in its 
excited state and the other is assumed to resides in its ground 
state. In the second, both atoms are assumed to reside in their 
excited states.

\subsubsection{Initial state of only one atom excited: 
$|\Phi_{0}\rangle =|e_{1}\rangle |g_{2}\rangle$.}

\begin{figure}[ht]
   \begin{center}
    \mbox{ \psfig{file=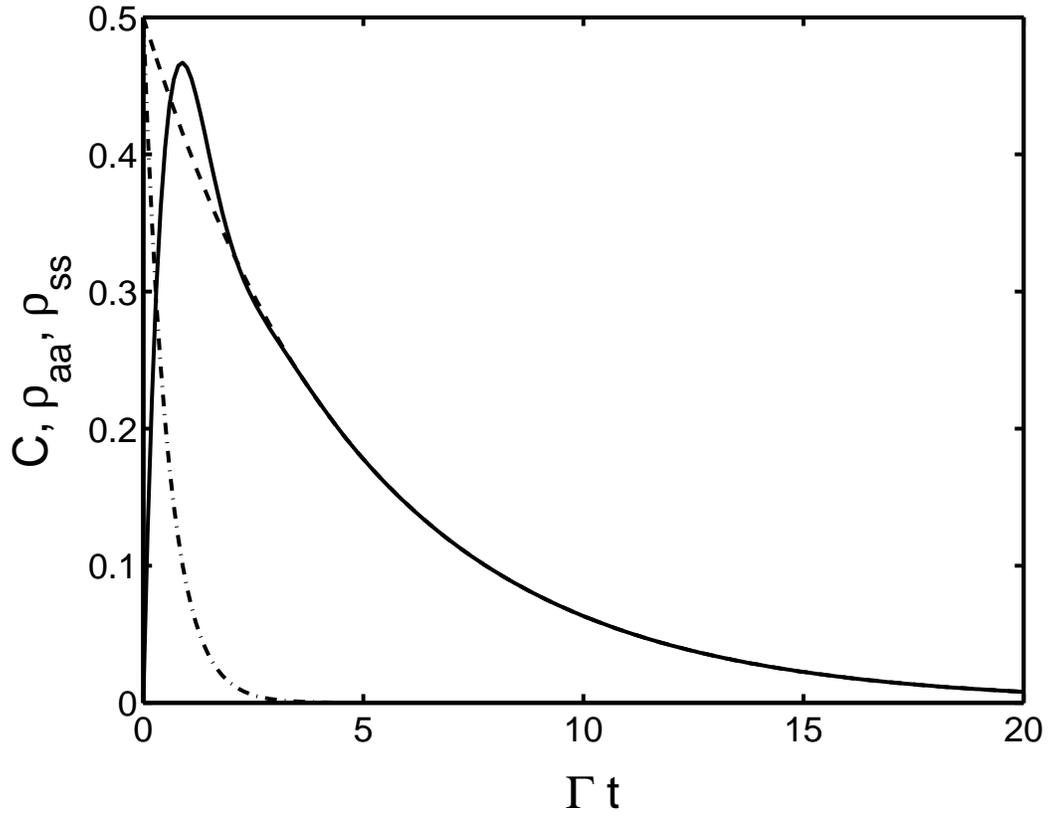,width=14cm}}
      \end{center}
\caption{Time evolution of the concurrence $C$ (solid line) and the
population of the antisymmetric state $\rho_{aa}(t)$ (dashed line) and 
the symmetric state $\rho_{ss}(t)$ (dashed-dotted line) for the atoms 
prepared initially in the unentangled state
$|\Phi_{0}\rangle =\left|e_{1}\right\rangle \left|g_{2}\right\rangle$, 
with $\bar{\mu}\perp \bar{r}_{12}$, and $r_{12}=\lambda/6$ 
$(\Gamma_{12}= 0.79\,\Gamma , \Omega_{12} =1.12\,\Gamma )$.}
    \label{fg1}
    \end{figure}

In the basis of the collective states of the system, the initial 
condition of only one atom excited corresponds to the initial 
condition with non-zero density matrix elements
$\rho_{ss}(0)=\rho_{aa}(0)=\rho_{sa}(0)=\rho_{as}(0)=1/2$.
In Fig.~\ref{fg1}, we plot the time evolution of the populations
$\rho_{ss}(t)$, $\rho_{aa}(t)$ and the concurrence $C$, which we have
found solving the equations of motion (\ref{eq6}) with the
condition that initially at $t=0$ one of the atoms was prepared in
its excited state and the other in the ground state.
One can see from Fig.~\ref{fg1} that the concurrence is zero at $t=0$; 
in other words there is no
entanglement in the system at $t=0$. The time evolution of the
concurrence reflects the time evolution of the entanglement between
the atoms. It is seen from Fig.~\ref{fg1} that the concurrence builds 
up as time
develops indicating that spontaneous emission can create entanglement
between the initially unentangled atoms. As time progresses, the
concurrence rapidly increases and reaches a maximum at time close to
the point where the symmetric state becomes depopulated.
At later times, the concurrence slowly decreases and overlaps with the
time evolution of the population of the antisymmetric state.

The physical understanding of the creation of the transient entanglement 
can be achieved by considering
the time-dependent density matrix of the system, which for the initial
condition of only one atom excited has the following form
\begin{eqnarray}
\rho (t) &=& \left(
\begin{array}{cccc}
0 & 0 & 0 & 0 \\
0 & \rho_{ss}(t) & \rho_{sa}(t) & 0 \\
0 & \rho_{as}(t) & \rho_{aa}(t) & 0 \\
0 & 0 & 0 & \rho_{gg}(t)
\end{array}
\right) \ , \label{eq8}
\end{eqnarray}
where the non-zero time-dependent density matrix elements are given by
\begin{eqnarray}
    \rho_{ss}(t) &=& \frac{1}{2}\exp \left[-\left(\Gamma 
    +\Gamma_{12}\right)t\right] \ ,\nonumber \\
    \rho_{aa}(t) &=& \frac{1}{2}\exp \left[-\left(\Gamma 
    -\Gamma_{12}\right)t\right] \ ,\nonumber \\
    \rho_{gg}(t) &=& 1- \exp \left(-\Gamma t\right)\cosh \Gamma_{12}t
    \ ,\nonumber \\
    \rho_{as}(t) &=& \rho_{sa}^{\ast}(t) =\frac{1}{2}\exp 
    \left[-\left(\Gamma 
    +2i\Omega_{12}\right)t\right] \ .\label{eq8a}
\end{eqnarray}

One can see that the density matrix is not diagonal due to the
presence of coherences $\rho_{sa}(t)$ and $\rho_{as}(t)$. The density
matrix can be rediagonalized to give new "diagonal" states
\begin{eqnarray}
\ket {\Psi_{1}} &=& \left\{\left[P_{1}(t)-\rho_{ss}(t)\right]\ket a +
\rho_{as}(t)\ket s
\right\}/\left\{\left[P_{1}(t)-\rho_{ss}(t)\right]^{2}
+\left|\rho_{as}(t)\right|^{2}\right\}^{\frac{1}{2}} \ ,\nonumber \\
\ket {\Psi_{2}} &=& \left\{\rho_{as}(t)\ket a +
\left[P_{2}(t)-\rho_{aa}(t)\right]\ket s
\right\}/\left\{\left[P_{2}(t)-\rho_{aa}(t)\right]^{2}
+\left|\rho_{as}(t)\right|^{2}\right\}^{\frac{1}{2}} \ ,\nonumber \\
\ket {\Psi_{3}} &=& \ket g \ ,\nonumber \\
\ket {\Psi_{4}} &=& \ket e \ ,\label{eq9}
\end{eqnarray}
where the diagonal probabilities (populations of the new states) are
\begin{eqnarray}
P_{1}(t) &=& \exp \left(-\Gamma t\right)\cosh \Gamma_{12}t \ ,\nonumber \\
P_{2}(t) &=& 0 \ ,\nonumber \\
P_{3}(t) &=& 1- \exp \left(-\Gamma t\right)\cosh \Gamma_{12}t \ ,\nonumber \\
P_{4}(t) &=& 0 \ .\label{eq10}
\end{eqnarray}
Thus, the coherences $\rho_{sa}(t)$ and $\rho_{as}(t)$ cause the
system to evolve effectively only between two states: the ground 
product state $ \ket g$ and the entangled state 
$\ket {\Psi_{1}}$, which is a linear combination of the states
$\ket s$ and $\ket a$. In this case, the density matrix of the system 
is diagonal for all times $t$, and is given by
\begin{eqnarray}
     \rho (t) = P_{1}(t)\ket {\Psi_{1}}\bra {\Psi_{1}}
     +\rho_{gg}(t)\ket g \bra g \ .\label{eq11}
\end{eqnarray}

It is easy to find from Eq.~(\ref{eq8a}) that
$|\rho_{as}(t)|^{2}=\rho_{aa}(t)\rho_{ss}(t)$, and then the state
$\ket {\Psi_{1}}$ can be written as
\begin{eqnarray}
\ket {\Psi_{1}} &=& \left(\sqrt{\rho_{ss}(t)}\ket s +
\sqrt{\rho_{aa}(t)}\ket a \right)/\sqrt{\rho_{aa}(t)+ \rho_{ss}(t)}
\ ,\label{eq12}
\end{eqnarray}
The state $\ket {\Psi_{1}}$ reduces to a nonentangled state
when the states $\ket s$ and $\ket a$ have equal populations.
On the other hand,
the state $\ket {\Psi_{1}}$ reduces to a maximally entangled state
when either $\rho_{ss}(t)$ or $\rho_{aa}(t)$ are equal to zero.

We are now in a position to understand quantitatively the method of 
creation of entanglement via spontaneous emission. The spontaneously 
induced entanglement   
results from unequal populations of the symmetric and antisymmetric 
states. This observation is supported by the Peres-Horodecki 
measure for entanglement. It is easy to show that the eigenvalues of 
the partial transposition of the density matrix (\ref{eq8}) are
\begin{eqnarray}
     \mu_{1} &=& \frac{1}{2}\left[\rho_{ss}(t) +\rho_{aa}(t) 
     -\rho_{as}(t)-\rho_{sa}(t)\right] \ ,\nonumber \\
     \mu_{2} &=& \frac{1}{2}\left[\rho_{ss}(t) +\rho_{aa}(t) 
     +\rho_{as}(t)+\rho_{sa}(t)\right] \ ,\nonumber \\
     \mu_{3} &=& \frac{1}{2}\{\rho_{gg}(t)+
     \left[\rho_{gg}^{2}(t)
     +\left(\rho_{ss}(t)-\rho_{aa}(t)\right)^{2} 
     -\left(\rho_{as}(t)-\rho_{sa}(t)\right)^{2}
     \right]^{\frac{1}{2}}\} \ ,\nonumber \\
     \mu_{3} &=& \frac{1}{2}\{\rho_{gg}(t)-
     \left[\rho_{gg}^{2}(t)
     +\left(\rho_{ss}(t)-\rho_{aa}(t)\right)^{2} 
     -\left(\rho_{as}(t)-\rho_{sa}(t)\right)^{2}
     \right]^{\frac{1}{2}}\} \ .\label{eq12a}
\end{eqnarray}
It is obvious from Eqs.~(\ref{eq12a}) and (\ref{eq8a}) that 
$\mu_{1}, \mu_{2}$ and $\mu_{3}$ are always positive.
The eigenvalue $\mu_{4}$ becomes negative if and only if
\begin{eqnarray}
     |\rho_{ss}(t)-\rho_{aa}(t)| >
     |\rho_{as}(t)-\rho_{sa}(t)| \ ,\label{eq12b}
\end{eqnarray}
which, according to Eq.~(\ref{eq8a}), is satisfied if 
$\rho_{ss}(t)\neq \rho_{aa}(t)$.  

Since
the population of the symmetric state decays faster than the
antisymmetric state (see Eq.~(\ref{eq8a})), at time when the state
$\ket s$ becomes depopulated, the state $\ket {\Psi_{1}}$ reduces to
the maximally entangled antisymmetric state $\ket a$.
The above analysis give clear evidence that a transient entanglement
created by spontaneous emission can appear only in spatially extended
two-atom systems where the antisymmetric
state fully participates in the dynamics of the system.

\subsubsection{Initial state of both atoms excited: 
$|\Phi_{0}\rangle =|e_{1}\rangle |e_{2}\rangle$.}

\begin{figure}[h]
   \begin{center}
     \mbox{ \psfig{file=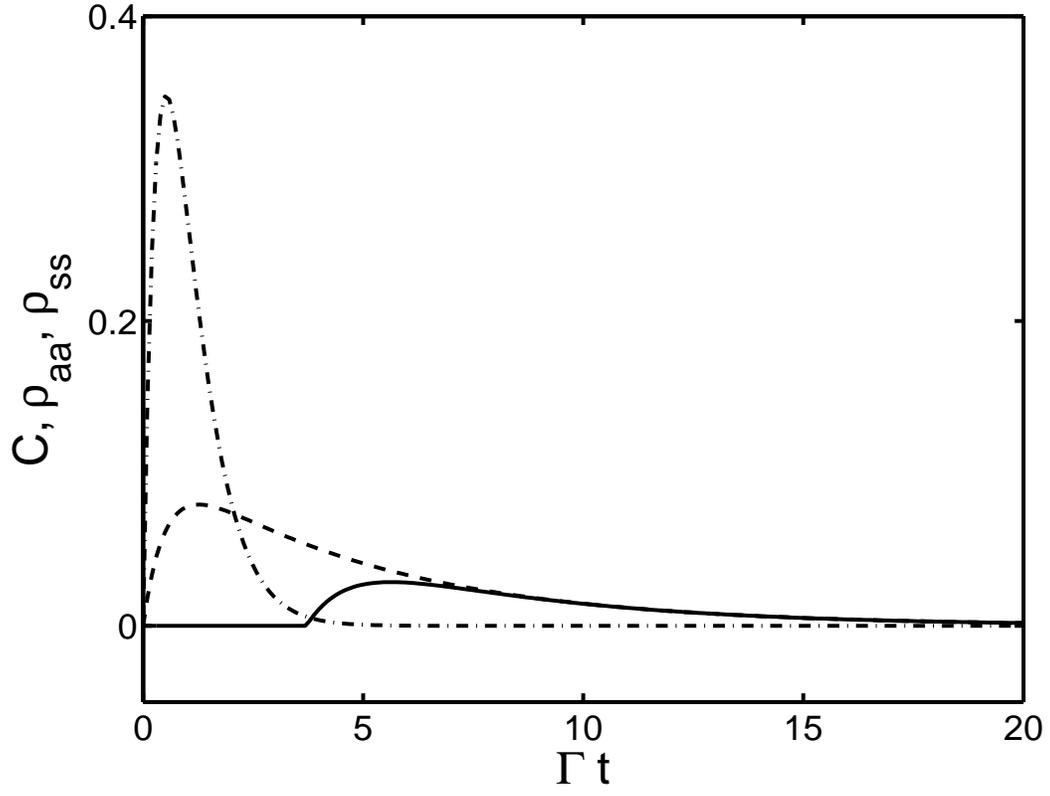,width=14cm}}
      \end{center}
\caption{Time evolution of the concurrence $C$ (solid line) and the
population of the antisymmetric state $\rho_{aa}(t)$ (dashed line) and 
the symmetric state $\rho_{ss}(t)$ (dashed-dotted line) for the initial 
unentangled state of both atoms excited,
$|\Phi_{0}\rangle =\left|e_{1}\right\rangle \left|e_{2}\right\rangle$, 
with $\bar{\mu}\perp \bar{r}_{12}$, and $r_{12}=\lambda/6$ 
$(\Gamma_{12}= 0.79\, \Gamma , \Omega_{12} =1.12\, \Gamma )$.}
    \label{fg2}
    \end{figure}

The role of the antisymmetric state in the entanglement creation is
more evident when we choose the initial state of both atoms in their
excited states. Here, $\rho_{ee}(0)=1$ and the initial
values of the remaining density matrix elements are zero. Note, that 
in this case there are no any initial coherences between the atoms, 
and also there are no any initial coherences between the collective 
states. Recent analysis of the entanglement creation in
the Dicke model have shown that there is no entanglement creation in
spontaneous emission when the atoms are initially prepared in their
excited states~\cite{klak,bash}. In Fig.~\ref{fg2}, we show the 
concurrence $C$ and the populations $\rho_{ss}(t)$
and $\rho_{aa}(t)$ for the case of two spatially separated atoms
prepared in their excited states at $t=0$. Since, at
$t=0$, only the upper state $\ket e$ is occupied, one finds from
Eq.~(\ref{eq6}) that the spontaneous emission populates the symmetric
and antisymmetric states with different rates. Thus, at early times
the symmetric state is more populated than the antisymmetric state.
Again, there is no entanglement between the atoms at $t=0$ as the
initial state $\ket e$ is an unentangled state. However, the figure
clearly demonstrates that contrary to the Dicke model, a transient
entanglement can be generated in spontaneous emission from two
spatially separated atoms. Note that the entanglement is not
generated until the population $\rho_{ss}(t)$ becomes zero. 

This effect can be explained by the Peres-Horodecki measure for 
entanglement.
Consider the time evolution of the density operator of the system, with
the initial condition of $\rho_{ee}(0)=1$, has the diagonal form for
all times
\begin{eqnarray}
     \rho (t) = \rho_{ee}(t)\ket e \bra e
     +\rho_{ss}(t)\ket s \bra s
     +\rho_{aa}(t)\ket a \bra a
     +\rho_{gg}(t)\ket g \bra g \ ,\label{eq13}
\end{eqnarray}
where $\rho_{ii}(t)$ $(i=e,s,a,g)$ are the time dependent populations
of the collective atomic states. According to the Peres-Horodecki
criterion for entanglement, the two-atom system represented by the 
density matrix (\ref{eq13}) is entangled when
\begin{eqnarray}
|\rho_{ss}(t) -\rho_{aa}(t)|>2\sqrt{\rho_{ee}(t)\rho_{gg}(t)} \ 
.\label{eq13a}
\end{eqnarray}
Thus, in contrast to the case of 
only one atom excited, it is not enough to produce an unbalanced 
population distribution between the states $|s\rangle$ and 
$|a\rangle$ to obtain an entanglement in the system. The reason is 
the presence of a population in the unentangled state $|e\rangle$. In 
order to analyse the dependence of the entanglement on the populations 
of the collective states, we solve Eq.~(\ref{eq6}) with the initial 
condition $\rho_{ee}(0)=1$ and find that the time-dependent 
populations are given by
\begin{eqnarray}
    \rho_{ee}(t) &=& \exp (-2\Gamma t) \ ,\nonumber \\
    \rho_{ss}(t) &=& \frac{\Gamma +\Gamma_{12}}{\Gamma -\Gamma_{12}} 
    \left\{ \exp \left[-\left(\Gamma +\Gamma_{12}\right)t\right] 
    -\exp \left(-2\Gamma t\right)\right\} \ ,\nonumber \\
    \rho_{aa}(t) &=& \frac{\Gamma -\Gamma_{12}}{\Gamma +\Gamma_{12}} 
    \left\{ \exp \left[-\left(\Gamma -\Gamma_{12}\right)t\right] 
    -\exp \left(-2\Gamma t\right)\right\} \ .\label{eq13b}
\end{eqnarray}
One can see from Eq.~(\ref{eq13b}) that in general the populations 
decay with different decay rates. However, for small interatomic 
separations, $\Gamma_{12}\approx \Gamma$, and then the upper state 
$|e\rangle$ and the symmetric state $|s\rangle$ decay with the same 
rate $(2\Gamma)$, whereas the antisymmetric state decays with a 
significantly reduced rate $\Gamma -\Gamma_{12}$. 
At early times the population is mostly in the state $|e\rangle$, and 
then the inequality (\ref{eq13a}) is not satisfied. The inequality 
(\ref{eq13a}) is not satisfied until $\rho_{ee}(t)\approx 0$. Since 
the population of the symmetric state decays with the same rate as 
$\rho_{ee}(t)$, at time where $\rho_{ee}(t)\approx 0$ the population 
$\rho_{ss}(t)\approx 0$.
Thus, the lack of the entanglement for $\rho_{ss}(t)\neq 0$ can be 
attributed to a large population of the product state $|e\rangle$ which, 
in turn, decays on the same time scale as $\rho_{ss}(t)$.

At time $t_{s}$ when $\rho_{ss}(t)=0$, the upper state $\ket e$ is
also depopulated, but there is still some population accumulated in the
antisymmetric state, as a result of a slow decay of the population with
the reduced rate $(\Gamma -\Gamma_{12})$. Therefore, for times larger
than $t_{s}$, the system behaves effectively as a two-level system,
whose the density matrix can be written as
\begin{eqnarray}
     \rho (t) = \rho_{aa}(t)\ket a \bra a
     +\rho_{gg}(t)\ket g \bra g \ .\label{eq14}
\end{eqnarray}
Following the Peres-Horodecki measure for entanglement, we see that
the atoms are entangled until the antisymmetric state is populated,
i.e. when $\rho_{aa}(t)\neq 0$.
The entanglement persists for a longer time and slowly decays to zero
on a time scale of order $\sim (\Gamma
-\Gamma_{12})^{-1}$, which for small interatomic separations is much
longer than the single atom decay rate $\Gamma^{-1}$.

\subsection{Non-identical atoms}

Although most of the work on entanglement is concerned with identical
atoms, there can be interesting information available on entanglement
creation with non-identical atoms. In this section, we give 
illustrative examples of both the entanglement creation and 
entanglement transfer between two nonidentical atoms with different 
transition frequencies.

\begin{figure}[h]
   \begin{center}
          \mbox{ \psfig{file=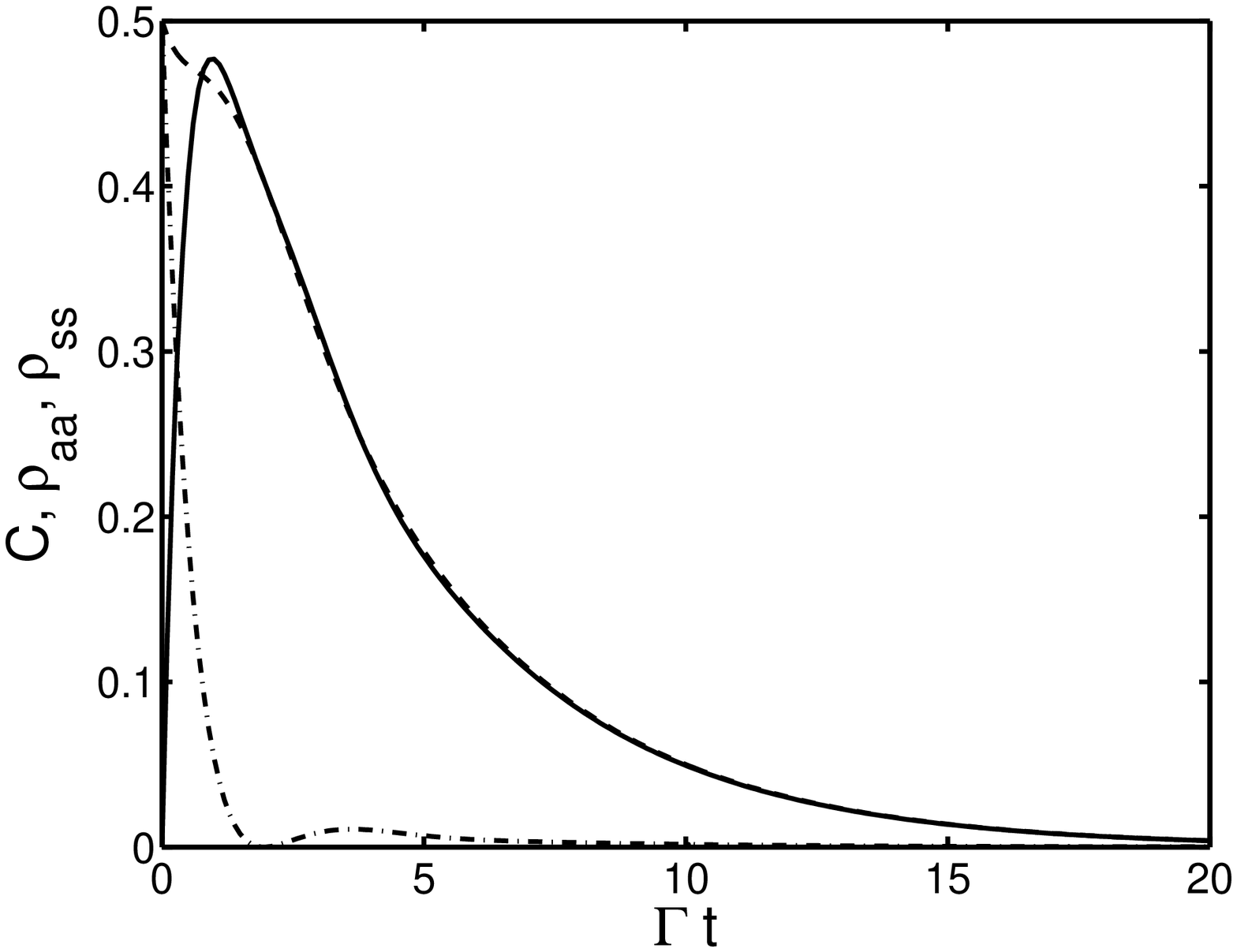,width=14cm}}
      \end{center}
\caption{Time evolution of the concurrence $C$ (solid line) and the
population of the antisymmetric state $\rho_{aa}(t)$ (dashed line) and 
the symmetric state $\rho_{ss}(t)$ (dashed-dotted line) for two 
non-identical atoms with $\Delta
=\Gamma$, $\bar{\mu}\perp \bar{r}_{12}$, and $r_{12}=\lambda/6$ 
$(\Gamma_{12}= 0.79\,\Gamma , \Omega_{12} =1.12\,\Gamma )$. The atoms were 
initially in the unentangled state
$|\Phi_{0}\rangle =\left|e_{1}\right\rangle \left|g_{2}\right\rangle$.}
    \label{fg3}
    \end{figure}

\begin{figure}[h]
   \begin{center}
          \mbox{ \psfig{file=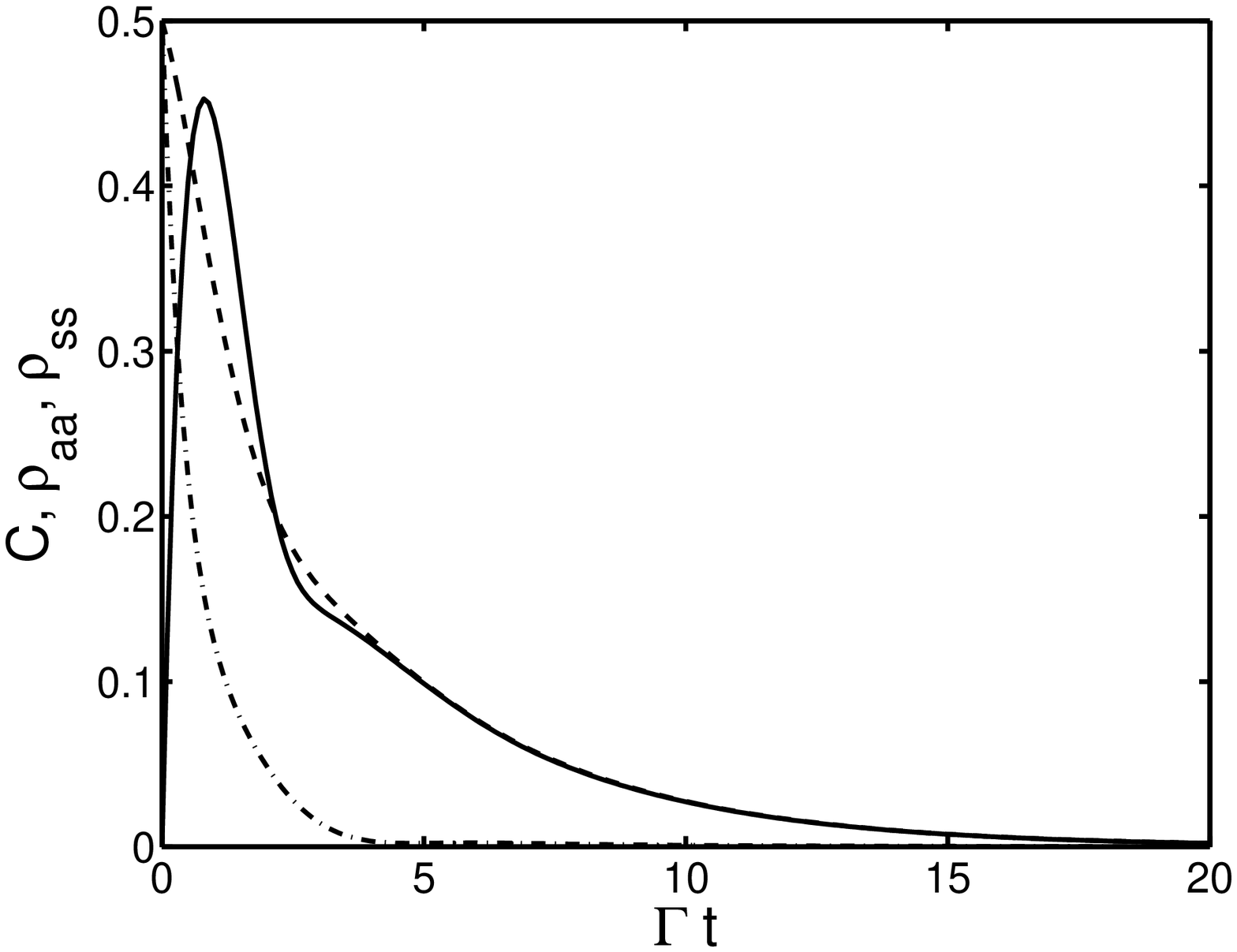,width=14cm}}
      \end{center}
\caption{Time evolution of the concurrence $C$ (solid line) and the
population of the antisymmetric state $\rho_{aa}(t)$ (dashed line) and 
the symmetric state $\rho_{ss}(t)$
(dashed-dotted line) for two non-identical atoms with $\Delta
=\Gamma$, $\bar{\mu}\perp \bar{r}_{12}$, and $r_{12}=\lambda/6$
$(\Gamma_{12}= 0.79\,\Gamma , \Omega_{12} =1.12\,\Gamma )$. The
atoms were initially in the unentangled state
$|\Phi_{0}\rangle =\left|g_{1}\right\rangle \left|e_{2}\right\rangle$.}
    \label{fg4}
    \end{figure}

Figure~\ref{fg3} shows the concurrence $C$ and the populations 
$\rho_{ss}(t)$ and $\rho_{aa}(t)$ as a
function of time for the initial condition of only one atom excited.
We take $\Delta =\Gamma$ that the atom $"2"$ has a higher frequency
(energy) than the atom $"1"$, i.e. $\omega_{2}>\omega_{1}$, and
assume that initially the atom of the higher transition frequency
was in its excited state while the other atom of lower transition
frequency was in the ground state. As in the case of identical atoms,
there is no initial entanglement between the atoms, and at
early times the entanglement builds up rapidly to a maximum appearing
at short time $\Gamma t<1$. However, comparing with the entanglement
for identical atoms, shown in Fig.~\ref{fg1}, we see that the maximum 
of the entanglement obtained with nonidentical atoms is greater than that
obtained with identical atoms. It is interesting to note that after
passing through the maximum, the entanglement decays with two different
time scales. This effect is more pronounced if we choose the atom of
the lower frequency to be initially in its excited state and
the other atom of higher frequency in the ground state. We illustrate
this in Fig.~\ref{fg4}, where we plot the concurrence $C$ and the populations
$\rho_{ss}(t)$ and $\rho_{aa}(t)$ as a function of time for the same
parameters as in Fig.~\ref{fg3}. The two decay time scales are well 
resolved. At early times the entanglement decays on a
time scale of order $\Gamma^{-1}$, and next the time decay "jumps"
into the time scale of order $(\Gamma -\Gamma_{12})^{-1}$, which
coincides with the decay of the population of the antisymmetric
state.

The enhanced entanglement and the two different decay time
scales can be explained as a consequence of the coherent transfer of the
population from the symmetric to the antisymmetric state, that is
absent for identical atoms. In order to show this, consider the
dynamics of the entangled states $\ket s$ and $\ket a$, described by
Eq.~(\ref{eq6}), that in the case of initially only one atom
excited simplify to the following equations of motion
\begin{eqnarray}
      \dot{\rho}_{ss} &=& -\left(\Gamma
      +\Gamma_{12}\right)\rho_{ss}
      +i\Delta \left(\rho_{as}-\rho_{sa}\right) \ ,\nonumber \\
      \dot{\rho}_{aa} &=& -\left(\Gamma
      -\Gamma_{12}\right)\rho_{aa}
      -i\Delta \left(\rho_{as}-\rho_{sa}\right) \ ,\nonumber \\
      \dot{\rho}_{as} &=& -\left(\Gamma +2i\Omega_{12}\right)\rho_{as}
      +i\Delta \left(\rho_{ss}-\rho_{aa}\right) \ .\label{eq15}
\end{eqnarray}
Equations~(\ref{eq13}) are formally identical to the optical Bloch
equations for
a two-level atom driven by a coherent field~\cite{ebe}. Here, the detuning
$\Delta$ plays the role of a Rabi frequency that coherently
transfers the population between the symmetric and antisymmetric states.
Note that the interaction between the states does not involve the
ground state $\ket g$, and therefore is not accompanied by spontaneous
emission. Thus, the coherent transfer of the population between the
states is a decoherence free process. In this process, the
population is efficiently transferred from the more populated
symmetric state to the antisymmetric state before it decays to the
ground state leading to the enhancement of the entanglement.

\begin{figure}[h]
   \begin{center}
          \mbox{ \psfig{file=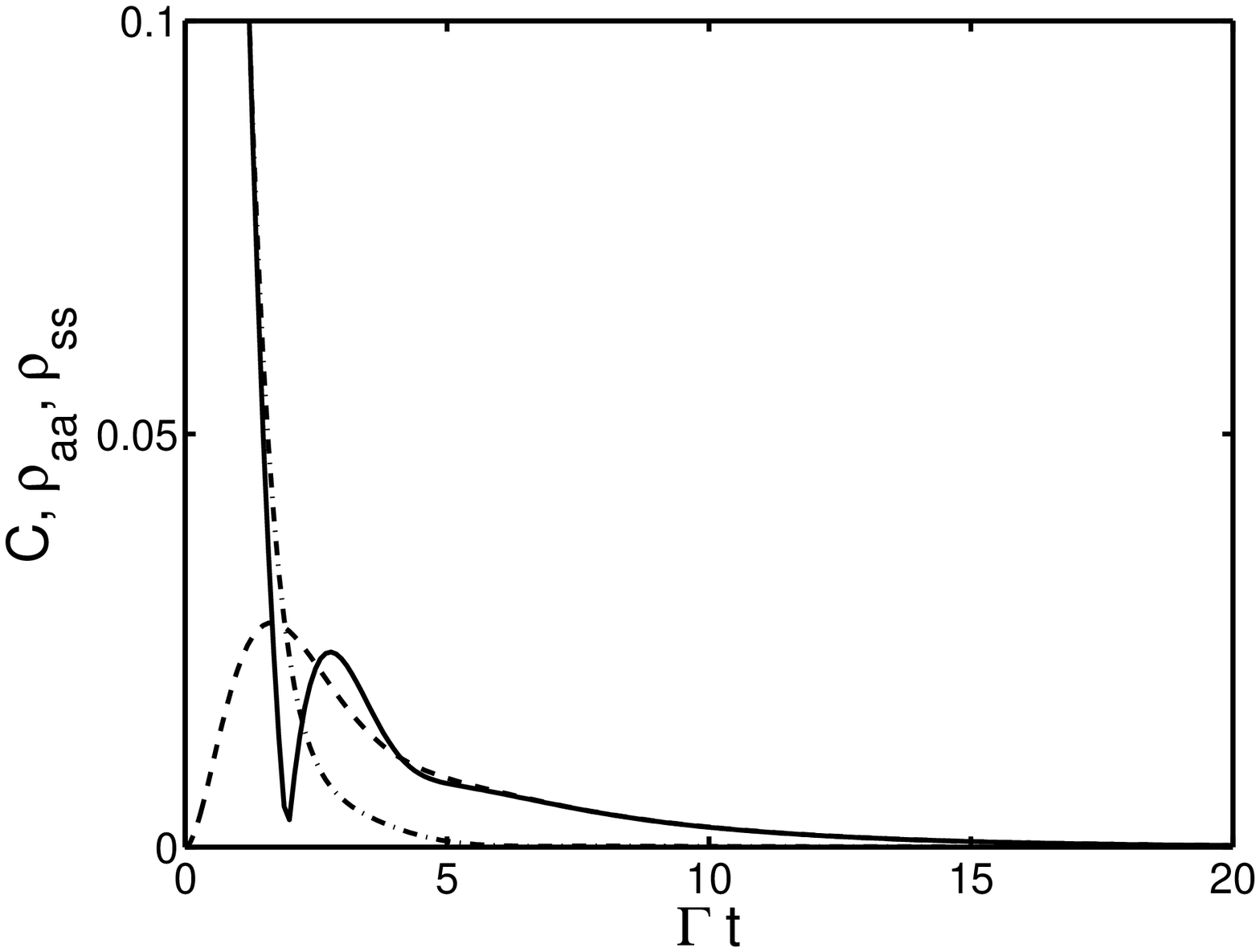,width=14cm}}
      \end{center}
\caption{Time evolution of the concurrence $C$ (solid line) and the
population of the antisymmetric state $\rho_{aa}(t)$ (dashed line) and the 
symmetric state $\rho_{ss}(t)$ 
(dashed-dotted line) for two non-identical atoms with $\Delta
=\Gamma$, $\bar{\mu}\perp \bar{r}_{12}$, and $r_{12}=\lambda/6$
$(\Gamma_{12}= 0.79\, \Gamma , \Omega_{12} =1.12\, \Gamma )$. The atoms 
were initially in the maximally
entangled state $|\Phi_{0}\rangle =\left|s\right\rangle$. For $t=0$ both
$\rho_{ss}$ and concurrence start from unity, and the figure is cut to
better visualize concurrence revival.}
    \label{fg5}
    \end{figure}

Finally, we point out one more difference between the entanglement
creation in identical and nonidentical atoms. Figure~\ref{fg5} shows 
the concurrence $C$ and the populations $\rho_{ss}(t)$ and $\rho_{aa}(t)$ 
for the same parameters as in Fig.~\ref{fg3}, but with the
new initial condition $\rho_{ss}(0)=1$. In this case there is
perfect $(C=1)$ initial entanglement between the atoms. Since, at
$t=0$, only the state $\ket s$ is occupied, one finds that at early
times the initial entanglement decays with the enhanced rate
$(\Gamma +\Gamma_{12})$
corresponding to the decay rate of the population of the symmetric
state. The entanglement decays in time until the atoms become
disentangled. This happens at time $\Gamma t\approx 2$, where the
population is equally distributed over the entangled states,
$\rho_{ss}(t)=\rho_{aa}(t)$. However, as time develops, the
entanglement emerges again. This is remarkable as spontaneous 
emission is essentially an irreversible process, and one might expect
that spontaneous emission merely degrades the initial entanglement.
The revival effect is absent for 
identical atoms, and is due to the coherent transfer of the 
population from the symmetric to the antisymmetric state. The revival
of the entanglement at time $\Gamma t \approx 2$ is more easily
understood by reference to the density matrix of the system. It is 
seen from Fig.~\ref{fg5} that for times $\Gamma t > 2$ the population of the 
symmetric state is negligible, and therefore the system behaves as 
a two-level system whose the density matrix is in the diagonal form
$\rho (t) = \rho_{aa}(t)\ket a \bra a +\rho_{gg}(t)\ket g \bra g$.
Since $\rho_{aa}(t)\neq 0$ for $\Gamma t > 2$, the atoms are 
entangled until the population $\rho_{aa}(t)$ decays eventually 
to the ground state.

\begin{figure}[h]
   \begin{center}
          \mbox{ \psfig{file=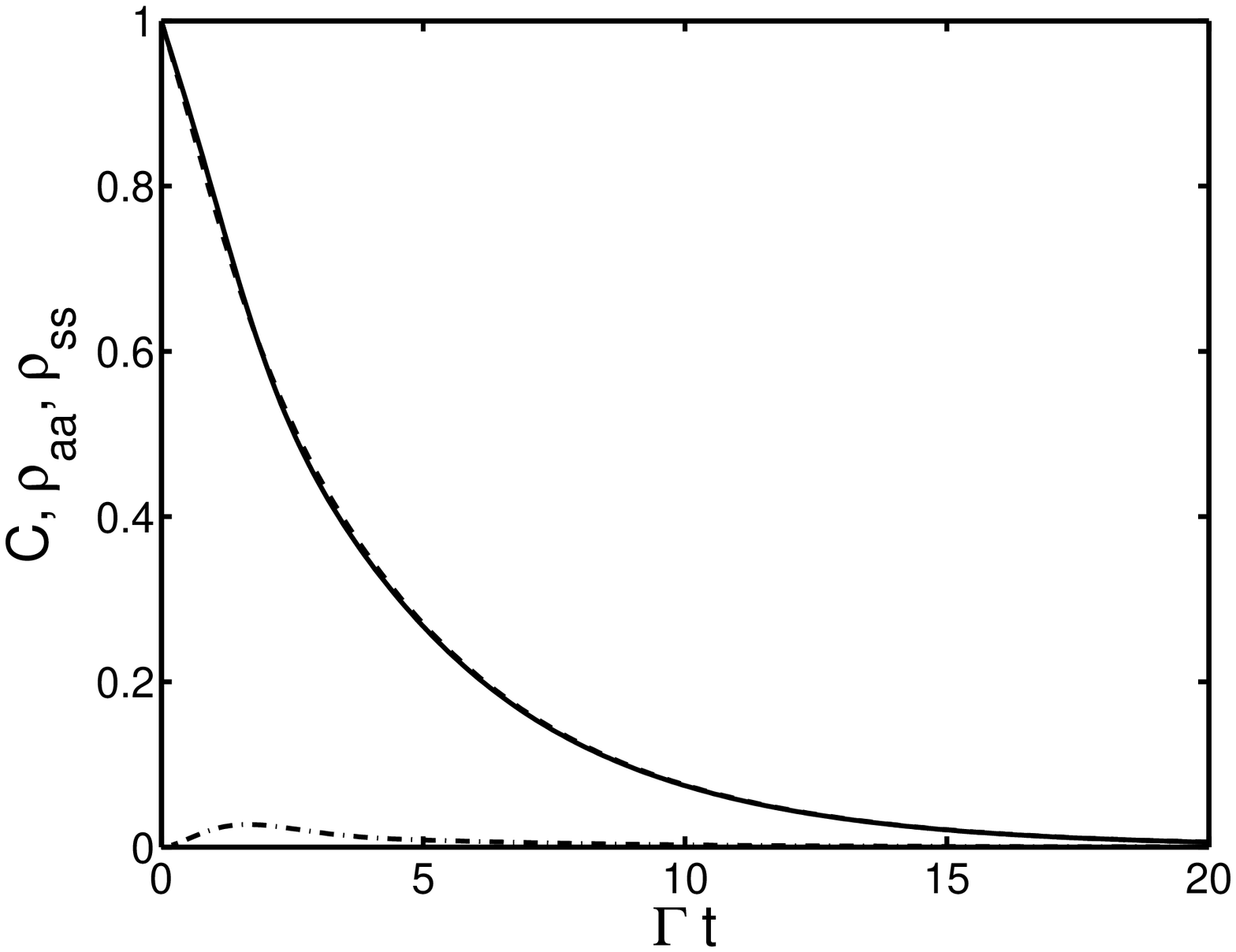,width=14cm}}
      \end{center}
\caption{Time evolution of the concurrence $C$ (solid line) and the
population of the antisymmetric state $\rho_{aa}(t)$ (dashed line) and 
the symmetric state $\rho_{ss}(t)$ 
(dashed-dotted line) for two non-identical atoms with $\Delta
=\Gamma$, $\bar{\mu}\perp \bar{r}_{12}$, and $r_{12}=\lambda/6$
$(\Gamma_{12}= 0.79\, \Gamma , \Omega_{12} =1.12\, \Gamma )$. 
The atoms were initially in the maximally
entangled state $|\Phi_{0}\rangle =\left|a\right\rangle$.}
    \label{fg6}
    \end{figure}

In contrast, if the atoms were initially prepared in the
antisymmetric state, $\rho_{aa}(0)=1$, the initial entanglement
remains in the antisymmetric state for all times $t$ even if a part of
the population is transferred to the symmetric state. This feature is
shown in Fig.~\ref{fg6}, where we plot the time evolution of the
concurrence and the population $\rho_{aa}(t)$ for the same parameters 
as in Fig.~\ref{fg5}, but with the initial condition $\rho_{aa}(0)=1$. 
It is clear from Fig.~\ref{fg6}
that the entanglement decays with the reduced rate $(\Gamma
-\Gamma_{12})$ corresponding to the decay rate of the population of
the antisymmetric state, and the coherent coupling does not transfer
much of the entanglement to the symmetric state.

Our calculations clearly demonstrate that the transient entanglement
induced from initially unentangled atoms depends crucially
on the presence of the antisymmetric state, that is characteristic
of spatially extended atomic systems.

\section{Summary}

In this paper, we have analysed the role of the antisymmetric state in 
entanglement creation by spontaneous emission from two spatially 
separated atoms. The results show that the generation of the  
entanglement is strongly dependent upon the population distribution 
between the symmetric and antisymmetric states of the system.
The entanglement is maximal when the population of the symmetric
state becomes zero. Thus, our results show the participation of the 
symmetric state in the atomic dynamics has a destructive effect on 
the entanglement. We have also considered the entanglement creation
in two non-identical atoms. We have found that the entanglement
can be enhanced by the process of the coherent transfer of the
population between the symmetric and antisymmetric states.
In addition, we have shown that the entanglement can decay with two
different time scales and exhibits a revival behavior 
due to the coherent transfer of the population
from the rapidly decaying symmetric to the slowly decaying
antisymmetric state before it decays to zero. Although the 
entanglement created by spontaneous emission appears only in the 
transient regime, it provides an useful information about 
entanglement sharing and entanglement transfer between two entangled 
states.

%\section*{References}

%\newpage

\end{document}